\documentclass[11pt]{article}
\usepackage{geometry}                
\usepackage{graphicx}
\usepackage{amssymb}
\usepackage{epstopdf}
\usepackage{subfigure}
\usepackage{hhline}
\usepackage{dcolumn}
\usepackage{multirow}
\usepackage{array}
\usepackage{color}
\usepackage{authblk}

\RequirePackage{lineno}
\title{Fragmentation in the $\phi^3$ Theory\\and the LPHD Hypothesis}
\date{}                                           

\author[1]{Karoly Urmossy\thanks{\texttt{e-mail: karoly.uermoessy@cern.ch}}}
\author[2]{Jan Rak}
\affil[1]{Wigner RCP of the HAS, 29-33 Konkoly-Thege Miklos Str., Budapest, Hungary, H-1121}
\affil[2]{University of Jyvaskyla, 9 Survontie Str., Jyvaskyla, Finland, FI-40014}

\newcommand{\be}{\begin{equation}}
\newcommand{\ee}[1]{\label{#1} \end{equation}}
\newcommand{\ba}{\begin{eqnarray}}
\newcommand{\ea}[1]{\label{#1} \end{eqnarray}}

\begin{document}
\maketitle


\begin{abstract}
We present analytic solution of the Dokshitzer--Gribov--Lipatov--Altarelli--Parisi (DGLAP) equation at leading order (LO) in the $\phi^3$ theory in 6 space-time dimensions. If the $\phi^3$ model was the theory of strong interactions, the obtained solution would describe the distribution of partons in a jet. We point out that the local parton-hadron duality (LPHD) conjecture does not work in this hypothetical situation. That is, treatment of hadronisation of shower partons is essential for the description of hadron distributions in jets stemming from proton-proton (pp) collisions at $\sqrt{s}$ = 7 TeV and from electron-positron ($e^+e^-$) annihilations at various collision energies. We use a statistical model for the description of hadronisation. 
\end{abstract}

\section{Introduction}
\label{sec:intro}

Recently, momentum fraction distributions of hadrons in jets stemming from electron-positron ($e^+e^-$) annihilations and proton-proton ($pp$) collisions have been described by simple analytic formulas obtained from statisitcal hadronisation models \cite{bib:UKeeFF,bib:UKeeFFs,bib:UKppFF}. The obtained fragmentation functions (FF) have succesfully been used in a perturbative quantum-chromodynamics (pQCD) improved parton model calculation to obtain the transverse momentum ($p_T$) spectrum of charged pions stemming from pp collisions at $\sqrt s$ = 7 TeV \cite{bib:Gribov80}, assuming $\log\log Q$ type scale evolution of the parameters of the FFs. This $Q$ scale dependence was conjectured based on fits of the newly proposed FFs to AKK-type \cite{bib:AKK} light-quark and gluon FFs. However, a solution of the Dokshitzer--Gribov--Lipatov--Altarelli--Parisi (DGLAP) \cite{bib:Dok,bib:Gri,bib:Alt} equations and a global fit to measured data is still missing in the case of these new FFs. Before addressing the full QCD problem, we examine the situation in the simplest asymptotically free field theory, the $\phi^3$  model in 6 spacetime dimensions, where there is only one type of parton. 

As we shall see, the \textit{local parton-hadron duality} (LPHD) hypothesis is not sufficient in the $\phi^3$ theory when trying to model strong interactions. In QCD, the distribution of partons produced in the branching process inside a jet describes the energy distribution of hadrons stemming from $e^+e^-$ annihilations (Sec.~7 in \cite{bib:Dok_pQCD}). However, in the $\phi^3$ model, parton branching does not produce enough soft 'gluons' for the description of the low energy regime of hadron distributions at parton level. At least that is what the solution of the DGLAP equation with leading-order (LO) splitting function (presented in Sec.~\ref{sec:Phi3}) suggests. To solve this problem, we introduce a statistical 'parton-hadronisation' function, and describe hadronic momentum fraction distributions in jets produced in $e^+e^-$ and pp collisions in Sec.~\ref{sec:Hadr}.

\section{Distribution of Partons in a Jet in the $\phi^3_6$ Theory}
\label{sec:Phi3}

In this section, we conjecture that a jet is initiated by an on-shell parton of momentum $P_{init} = (Q,0,0,Q)$, and obtain the longitudinal momentum fraction distribution $D(z,Q^2)$ of partons of momenta $p=zP_{init}$ in the jet from the DGLAP equation in the $\phi^3$ theory:
\be
\frac{d}{d\log Q^2} D\left(z,Q^2\right) \;=\; g^2 \int\limits_z^1  \frac{dy}{y} P(y) \, D\left(\frac{z}{y},Q^2\right) \;.
\ee{phi1}
At LO, the coupling $g^2(Q^2) = 1/\beta_0 \ln(Q^2/\Lambda^2)$, and the splitting function \cite{bib:Graz} is
\be
P(z) \;=\; z(1-z) - \frac{1}{12}\delta(1-z)\,.
\ee{phi2}
Eq.~(\ref{phi1}) factorizes in Mellin space:
\be
\frac{d}{d\log Q^2} \tilde{D}\left(s,Q^2\right) \;=\; g^2 \tilde{P}(s) \, \tilde{D}\left(s,Q^2\right)\,,
\ee{phi3}
where for a function $f$, 
\be
\tilde{f}(s) \;=\; \int\limits_0^1 dz\, z^{s-1} f(z)\,, \qquad f(z) \;=\; \frac{1}{2\pi}\int\limits_{-\infty}^\infty ds\, z^{-is} \tilde{f}(is)\,.
\ee{phi4}
The sollution of Eq.~(\ref{phi3}) with $t = \ln(Q^2/\Lambda^2)$ is
\be
\tilde{D}(s,t) \;=\; \tilde{D}(s,t_0)\, e^{\tilde{P}(s) \int\limits_{t_0}^t dt' g^2(t')} \;=\; \tilde{D}(s,t_0) \, e^{\tilde{P}(s)\, b(t)}\,,
\ee{phi5}
with
\be
\tilde{P}(s) \;=\; \frac{1}{(s+2)(s+1)} - \frac{1}{12}\,,\qquad b(Q^2) \;=\; \frac{1}{\beta_0}\ln\left[\frac{\ln(Q^2/\Lambda^2)}{\ln(Q^2_0/\Lambda^2)} \right]\,.
\ee{phi6}
As the initial parton had all of its own momentum, the initial parton distribution $D(z,Q^2_0) = \delta(z-1)$. Thus, from Eqs.~(\ref{phi4}--\ref{phi6}), the distribution of partons in the jet becomes
\be
 D(z,Q^2) \;\sim\; 
\delta(z-1) + \sum\limits_{k=1}^\infty \frac{b^k(Q^2)}{k!(k-1)!} \sum\limits_{j=0}^{k-1} \frac{(k-1+j)!}{j!(k-1-j)!} z \ln^{k-1-j}\left[\frac{1}{z}\right] \Big[ (-1)^j + (-1)^k z \Big]\,.
\ee{phi7}

As can be seen in Fig.~\ref{fig:D} (top panels), the distribution of partons in the jet Eq.~(\ref{phi7}) describes hadron momentum fraction distributions only in an intermediate range in the case of jets stemming from $e^+e^-$ and pp collisions. This effect might have been predicted from the terms in the gluon-to-gluon splitting function in QCD
\be
P_{gg}(z) \;\sim\; \frac{1-z}{z} + \frac{z}{1-z} + z(1-z) + c \,\delta(1-z)\,,
\ee{phi8}
not present in Eq.~(\ref{phi2}), and which enhance the production of very soft and very high-$z$ gluons.

\section{Introduction of a Hadronisation Function}
\label{sec:Hadr}

In this section, we introduce a hadronisation function $d(z)$ to describe the probability of a parton stemming from the branching proccess to produce some hadrons. This case, the hadron distribution in a jet becomes
\be
\frac{dN}{dz} \;=\;  \int\limits_z^1  \frac{dy}{y} D(y,Q^2) \, d\left(\frac{z}{y}\right)\;.
\ee{hadr1}
Eq.~(\ref{phi5}) with initial condition $D(z,Q^2_0) = \delta(z-1)$ (that is, $\tilde{D}(s,Q^2_0) = 1$) shows that the number of branchings in the parton evolution process has Poissonian distribution, as $\tilde{D} \sim \sum (\tilde{P}b)^k/k!$. In $z$-space, products of $\tilde{P}$-s are convolutions, thus Eq.~(\ref{hadr1}) can be written as
\be
\frac{dN}{dz} \;=\; \sum\limits_{k=0}^\infty \frac{b^k(Q^2)}{k!} \prod_{j=1}^k \int dy_j P(y_j) d(y_{k+1}) \delta(y_1 \cdots y_{k+1} - z)  \;.
\ee{hadr2}
If the splitting function $P(y)$ had a single peak at some $y^\ast$, Eq.~(\ref{hadr2}) could be approximated by
\be
\frac{dN}{dz} \;\approx\; \sum\limits_{k=0}^\infty \frac{\big[b(Q^2)\, P(y^\ast)\big]^k}{k! y^{\ast k}} d\left( \frac{z}{y^{\ast k}} \right)  \;.
\ee{hadr3}

To choose a simple model for the hadronisation function $d(z)$, we conjecture that this process is dominantly determined by the phasespace of the produced hadrons. Arguments supporting such a conjecture can be found in \cite{bib:BegunMic,bib:BegunVflukt,bib:Becattini1,bib:LiuWernerMic,bib:RybWlodMic}. Furtheremore, we assume that hadrons are collinear to their parent parton, and we neglect masses. Thisway, the momentum fraction distribution of one hadron out of $n$ hadrons stemming from the same parton becomes a one-dimensional microcanonical distribution \cite{bib:UKeeFF,bib:UKeeFFs,bib:UKppFF}:
\be
d_n(z) \;\sim\; \left(1 - z \right)^{n-2} \;.
\ee{hadr4}

\begin{figure}[h!]
\centering
\resizebox{0.4\columnwidth}{!}{\includegraphics{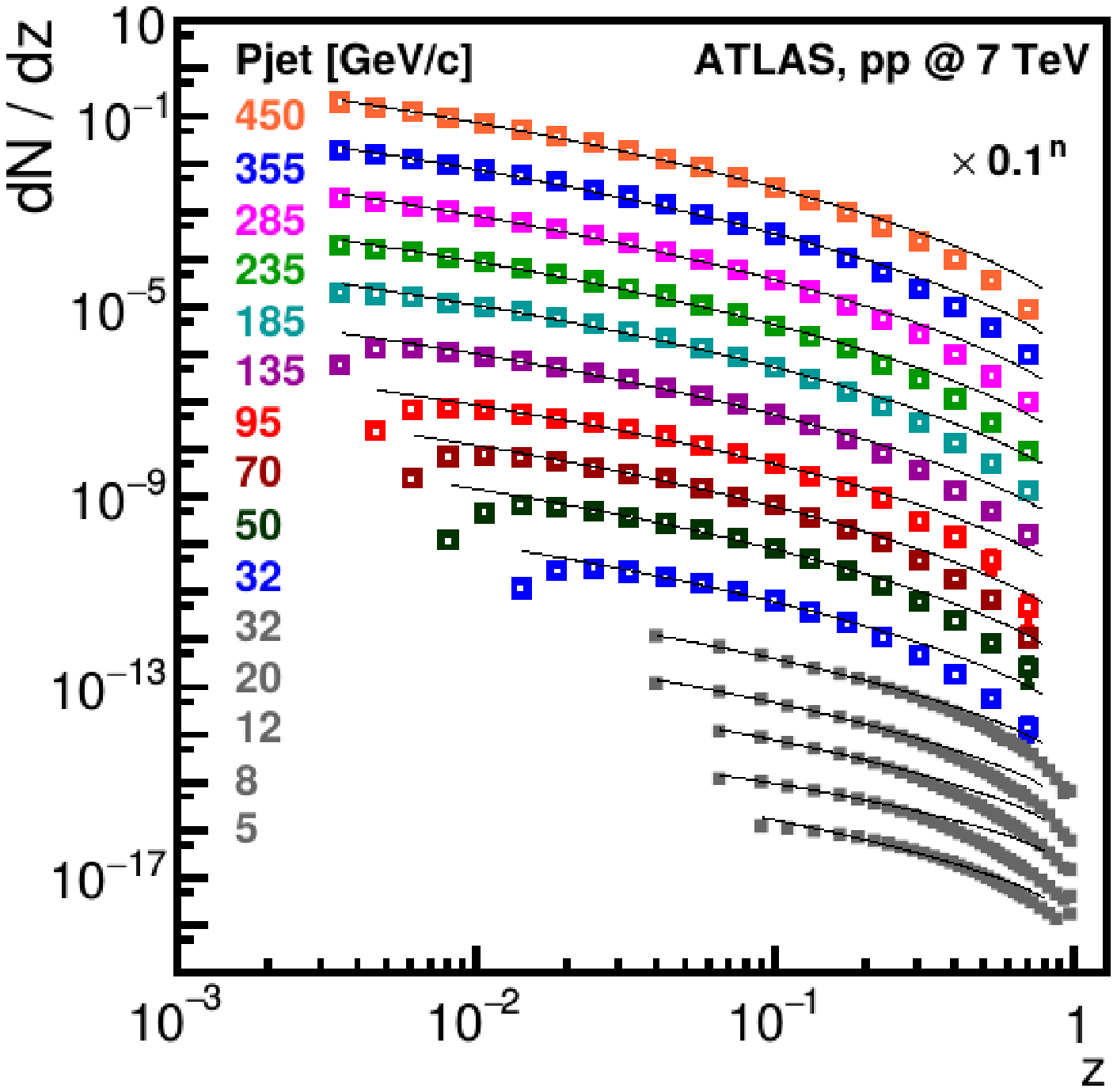}}
\resizebox{0.385\columnwidth}{!}{\includegraphics{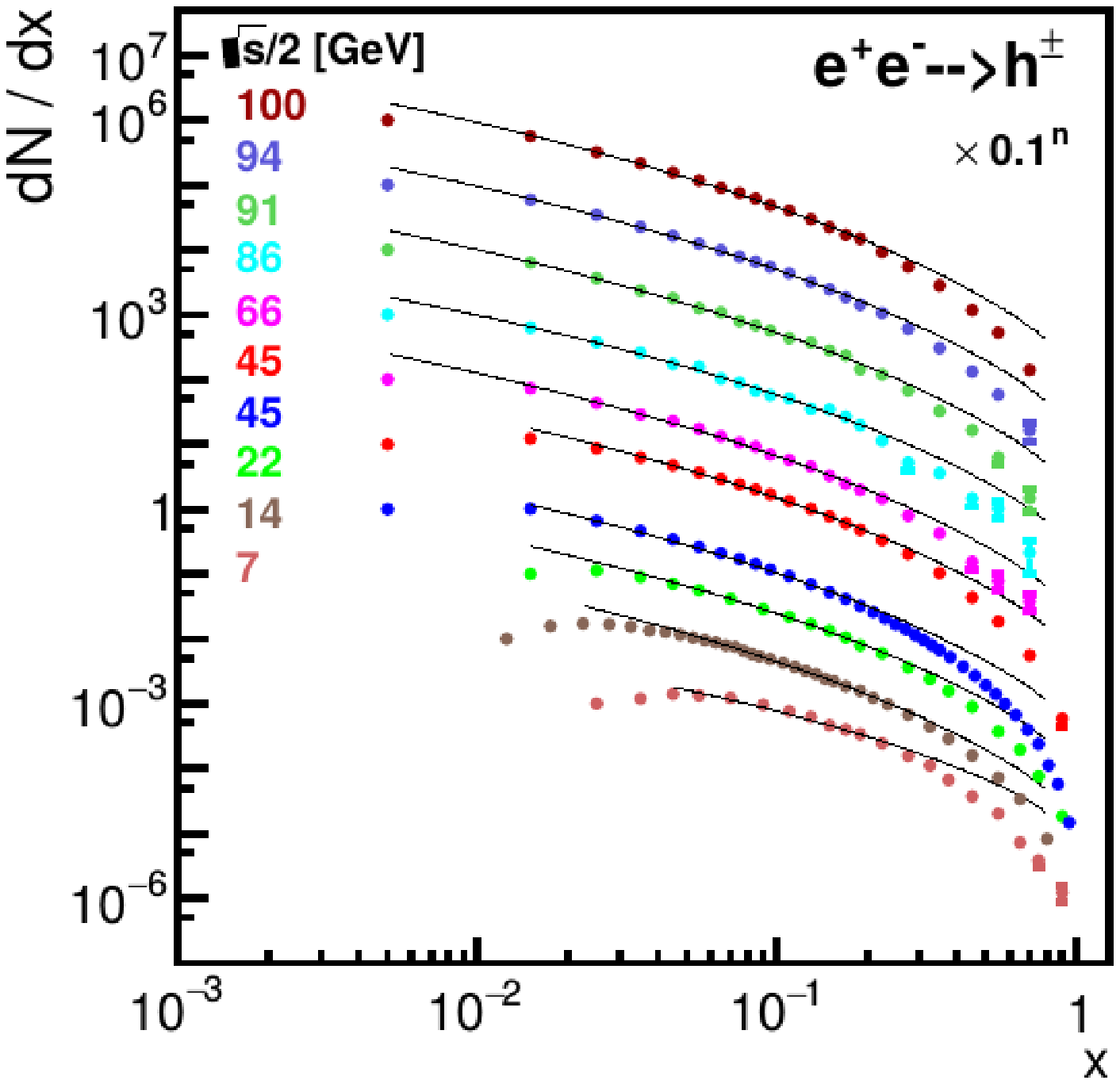}}
\resizebox{0.4\columnwidth}{!}{\includegraphics{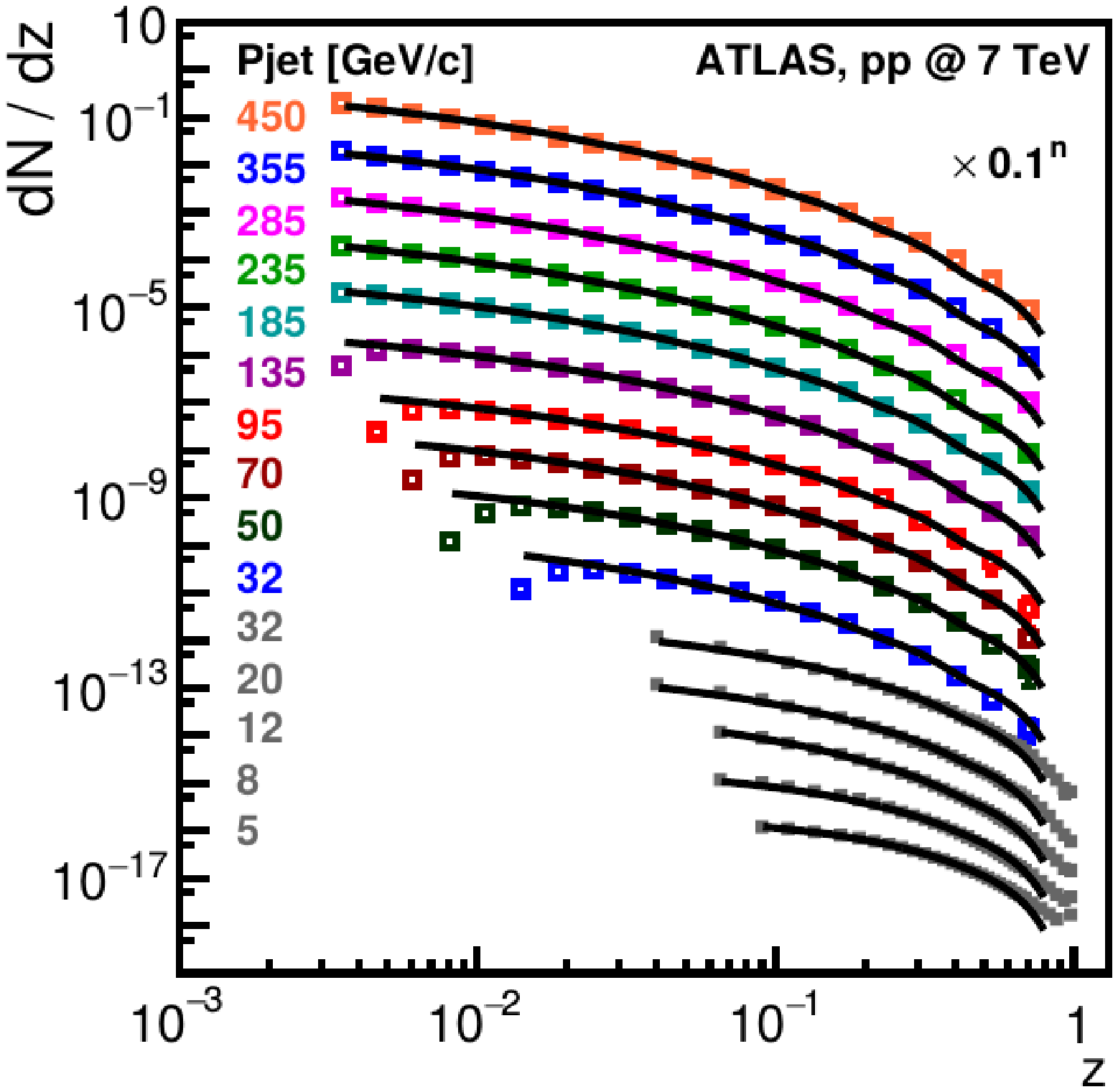}}
\resizebox{0.4\columnwidth}{!}{\includegraphics{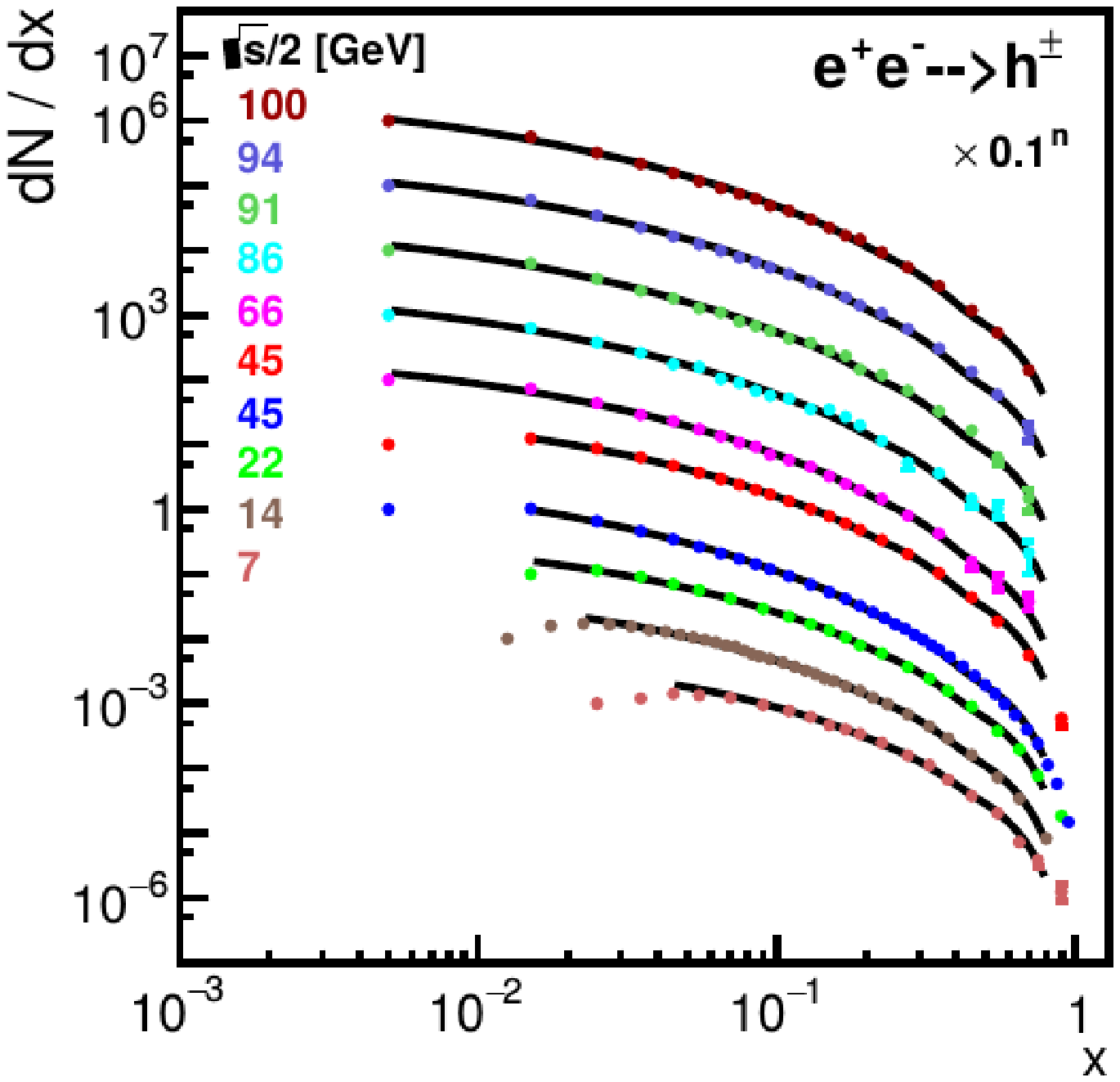}}
\resizebox{0.3\columnwidth}{!}{\includegraphics{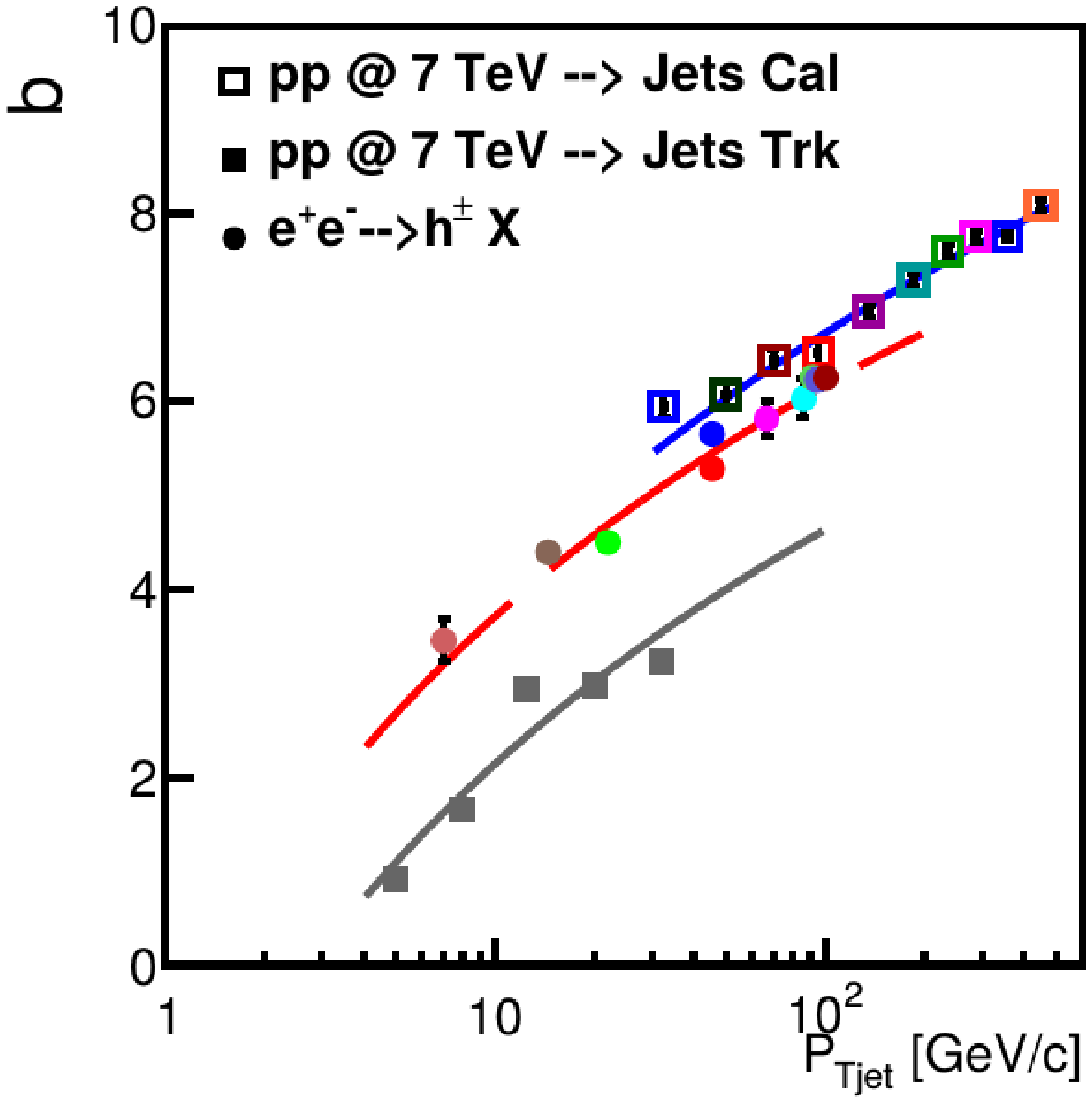}}
\caption{Momentum fraction distributions of charged hadrons in jets stemming from pp (\textbf{left}), and $e^+e^-$ (\textbf{right}) collisions. Data are compared to calculated results for partons Eq.~(\ref{phi7}) (\textbf{top}) and final state hadrons Eq.~(\ref{hadr5}) (\textbf{center}). Obtained values of the $b(Q^2)$ parameter are fitted with Eq.~(\ref{phi6}) (\textbf{bottom}). Data obtained by calorimetric \cite{bib:atlasFFpp7TeV} and track-based \cite{bib:atlasFFpp7TeV_low} jet reconstruction are represented by open and full rectangles in the case of pp collisions.\label{fig:D}}
\end{figure}

Though the maximum of the splitting function at LO in the $\phi^3$ theory (Eq.~(\ref{phi2})) at $y^\ast=1/2$ is not at all peak-like, we try out the approximation following from Eqs.~(\ref{hadr3},\ref{hadr4}):
\be
\frac{dN}{dz} \;\approx\; \sum\limits_{k=0}^\infty \frac{b^k(Q^2)}{k! 2^k } \left(1 -  2^k z \right)^{n-2}  \;.
\ee{hadr5}
Fig.~\ref{fig:D} (central panels) shows that Eq.~(\ref{hadr5}) provides a reasonably good fit of $dN/dz$ and $dN/dx$ distributions of hadrons in jets of various energy, stemming from pp \cite{bib:atlasFFpp7TeV,bib:atlasFFpp7TeV_low} and $e^+e^-$ \cite{bib:tassoX14,bib:tpcX29,bib:delphiX91,bib:opalX91,bib:opalX133,bib:opalX180,bib:opalX200} collisions. As expected, the model fails to reproduce hadron spectra at very low $z$, due to the lack of soft 'gluon' radiation in the $\phi^3$ theory. Besides, at $z\approx 1$, the curve of Eq.~(\ref{hadr5}) becomes uneven, as an artefact of the replacement of the continuous integral in Eq.~(\ref{hadr1}) by the integrand taken at $y\ast=1/2$.

Throughout the fitting, parameter $n=5$ has been used, while the obtained values of the $b(Q^2)$ are shown in Fig.~\ref{fig:D} (bottom panel) with $Q=\sqrt s$ for $e^+e^-$ and $Q=P_{jet}$ for pp collisions. The scale dependence of the $b$ parameter was fitted by the LO $\phi^3_6$ theory result Eq.~(\ref{phi6}) with $\Lambda = 0.2$ GeV (see Tab.~\ref{tab:bQ}). The values of $\beta_0$ and $Q_0$ coincide in case of jets stemming from $e^+e^-$ and pp collisions with calorimetric jet reconstruction \cite{bib:atlasFFpp7TeV}. However, these values differ in the case of pp data with track-based jet reconstruction \cite{bib:atlasFFpp7TeV_low}.
\begin{table}
\begin{center}
\begin{tabular}{ r|cc }
             & $\beta_0$       &  $Q_0$          \\[0.5mm] \hhline{===} \noalign{\smallskip}
  pp, Cal    & 0.168$\pm$0.008 &  1.484$\pm$0.171\\[0.5mm] \hline \noalign{\smallskip}  
  pp, Trk    & 0.185$\pm$0.007 &  2.775$\pm$0.125\\[0.5mm] \hline \noalign{\smallskip}
  $e^+e^-$         & 0.188$\pm$0.007 &  1.403$\pm$0.100\\[0.5mm] \hline \noalign{\smallskip}
\end{tabular}
\caption{Scale dependence of the $b(Q^2)$ parameter of Eq.~(\ref{hadr5}) obtained from fits shown in Fig.~\ref{fig:D}.\label{tab:bQ}}
\end{center}
\end{table}


\section*{Acknowledgement}
\label{sec:Conc}

This paper was also supported by the Hungarian OTKA Grant K104260.

\bibliographystyle{h-physrev3}
\bibliography{FFpaperBiblio}

\end{document}